\documentclass[aps,prl,superscriptaddress,floatfix,reprint]{revtex4-1}  

\usepackage{graphicx}
\usepackage{bm}
\usepackage{amssymb,amsmath}
\usepackage{float}
\usepackage{pstricks} 
\usepackage{epsfig,epsf,graphics}
\usepackage{color}
\usepackage{bbold}
\usepackage{xr} 

\newcounter{subfigure}[figure]

\newcommand{\bra}[1]{\left\langle #1 \right\vert}
\newcommand{\ket}[1]{\left\vert #1 \right\rangle}

\newcommand{\grad}[1]{\nabla }

\def\ket#1{ | #1 \rangle}
\def\bra#1{{\langle #1 |  }}
\newcommand{\ip}[2]{\langle #1 | #2 \rangle}
\newcommand{\adagg}{\hat{a}^{\dagger}}
\newcommand{\bdagg}{\hat{b}^{\dagger}}

\DeclareRobustCommand{\orderof}{\ensuremath{\mathcal{O}}}

\begin{document}

\title{Boson Sampling on a Photonic Chip}

\author{Justin B. Spring}
\email{j.spring1@physics.ox.ac.uk}
\affiliation{Clarendon Laboratory, Department of Physics, University of Oxford, OX1 3PU, UK}
\author{Benjamin J. Metcalf}
\affiliation{Clarendon Laboratory, Department of Physics, University of Oxford, OX1 3PU, UK}
\author{Peter C. Humphreys}
\affiliation{Clarendon Laboratory, Department of Physics, University of Oxford, OX1 3PU, UK}
\author{W. Steven Kolthammer}
\affiliation{Clarendon Laboratory, Department of Physics, University of Oxford, OX1 3PU, UK}
\author{Xian-Min Jin}
\affiliation{Clarendon Laboratory, Department of Physics, University of Oxford, OX1 3PU, UK}
\affiliation{Department of Physics, Shanghai Jiao Tong University, Shanghai 200240, PR China}
\author{Marco Barbieri}
\affiliation{Clarendon Laboratory, Department of Physics, University of Oxford, OX1 3PU, UK}
\author{Animesh Datta}
\affiliation{Clarendon Laboratory, Department of Physics, University of Oxford, OX1 3PU, UK}
\author{Nicholas Thomas-Peter}
\affiliation{Clarendon Laboratory, Department of Physics, University of Oxford, OX1 3PU, UK}
\author{Nathan K. Langford}
\affiliation{Clarendon Laboratory, Department of Physics, University of Oxford, OX1 3PU, UK}
\affiliation{Department of Physics, Royal Holloway, University of London, TW20 0EX, UK}
\author{Dmytro Kundys}
\affiliation{Optoelectronics Research Centre, University of Southampton, Southampton, SO17 1BJ, UK}
\author{James C. Gates}
\affiliation{Optoelectronics Research Centre, University of Southampton, Southampton, SO17 1BJ, UK}
\author{Brian J. Smith}
\affiliation{Clarendon Laboratory, Department of Physics, University of Oxford, OX1 3PU, UK}
\author{Peter G.R. Smith}
\affiliation{Optoelectronics Research Centre, University of Southampton, Southampton, SO17 1BJ, UK}
\author{Ian A. Walmsley}
\email{i.walmsley1@physics.ox.ac.uk}
\affiliation{Clarendon Laboratory, Department of Physics, University of Oxford, OX1 3PU, UK}

\begin{abstract}
While universal quantum computers ideally solve problems such as factoring integers exponentially more efficiently than classical machines, the formidable challenges in building such devices motivate the demonstration of simpler, problem-specific algorithms that still promise a quantum speedup.  We construct a quantum boson sampling machine (QBSM) to sample the output distribution resulting from the nonclassical interference of photons in an integrated photonic circuit, a problem thought to be exponentially hard to solve classically. Unlike universal quantum computation, boson sampling merely requires indistinguishable photons, linear state evolution, and detectors.  We benchmark our QBSM with three and four photons and analyze sources of sampling inaccuracy. Our studies pave the way to larger devices that could offer the first definitive quantum-enhanced computation.
\end{abstract}

\maketitle

Universal quantum computers require physical systems that are well-isolated from the decohering effects of their environment, while at the same time allowing precise manipulation during computation. They also require qubit-specific state initialization, measurement, and the generation of quantum correlations across the system~\cite{DiVincenzo2000,Raussendorf2001,Nielsen2003,Childs2009,Lovett2010}. Although there has been substantial progress in proof-of-principle demonstrations of quantum computation~\cite{walther05,lu07,lanyon07,lanyon11,lucero12}, simultaneously meeting these demands has proven difficult. This motivates the search for schemes that can demonstrate quantum-enhanced computation under more favorable experimental conditions. The power of one qubit~\cite{kl98}, permutational~\cite{j10} and instantaneous~\cite{sb09} quantum computation, are examples that have allowed investigation of the space between classical and universal quantum machines.

Boson sampling has recently been proposed as a specific quantum computation that is more efficient than its classical counterpart but only requires identical bosons, linear evolution, and measurement~\cite{Aaronson}.  The distribution of bosons that have passed through a linear system represented by the transformation $\mathbf{\Lambda}$ is thought to be exponentially hard to sample from classically~\cite{Aaronson}. The probability amplitude of obtaining a certain output is directly proportional to the permanent of a corresponding submatrix of $\mathbf{\Lambda}$.  The permanent expresses the wavefunction of identical bosons, which are symmetric under exchange~\cite{Caianiello1953,Troyansky1996}; in contrast, the Slater determinant expresses the wavefunction of identical fermions, which are antisymmetric under exchange.  While determinants can be evaluated efficiently, permanents have long been believed to be hard to compute~\cite{Valiant1979}; the best known algorithm scales exponentially with the size of the matrix~\cite{Ryser1963}.  One can envision a race between a classical and a quantum machine to sample the boson distribution given an input state and $\mathbf{\Lambda}$.  The classical machine would evaluate at least part of the probability distribution, which requires the computation of matrix permanents.  The QBSM instead creates indistinguishable bosons, directly implements $\mathbf{\Lambda}$, and records the outputs.  For a sufficiently large problem size, the quantum machine is expected to win~\cite{Aaronson}.

\begin{figure}[h!]
	\begin{center}
  \includegraphics[width=0.8\columnwidth]{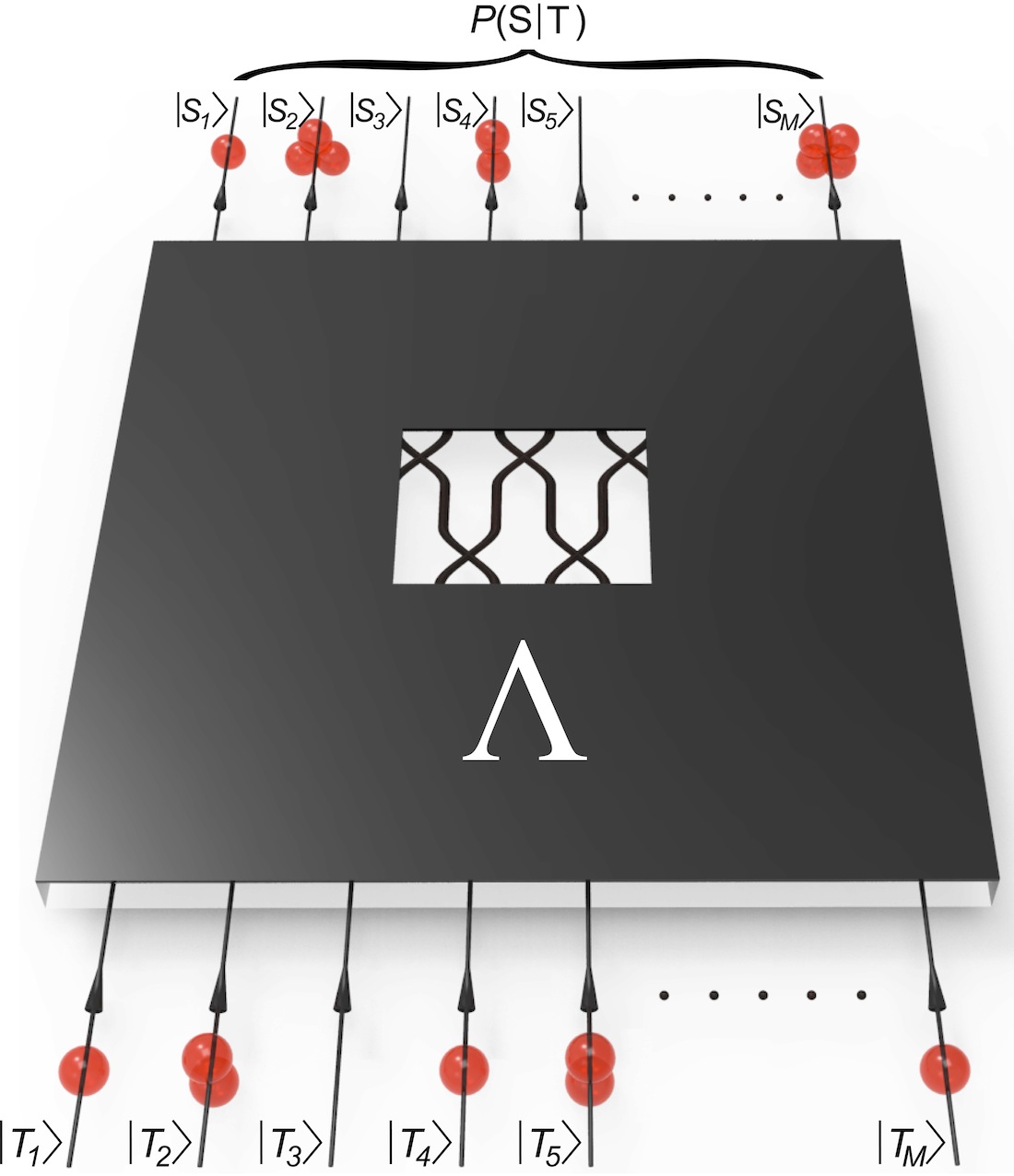}
  \caption{\textbf{Model of quantum boson sampling.}  Given a specified initial number state $\ket{\mathbf{T}} = \ket{T_1...T_M}$ and linear transformation $\mathbf{\Lambda}$, a quantum boson sampling machine efficiently samples from the distribution $P(\mathbf{S}|\mathbf{T})$ of possible outcomes $\ket{\mathbf{S}} = \ket{S_1...S_M}$.}
  \label{fig:concept}
  \end{center}
\end{figure}

Photonics is a natural platform to implement boson sampling since sources of indistinguishable photons are well-developed~\cite{Eisa2011}, and integrated optics offers a scalable route to large linear systems~\cite{Metcalf2012}.  In addition, photons have extremely low decoherence rates.  Importantly, boson sampling requires neither nonlinearities nor on-demand entanglement, unlike photonic approaches to universal quantum computation~\cite{Knill2001}. This clears the way for experimental boson sampling with existing photonic technology, building on the extensively studied two-photon Hong-Ou-Mandel (HOM) interference effect~\cite{Hong1987}.

A QBSM (Fig.~\ref{fig:concept}) samples the output distribution of a multi-particle bosonic quantum state $\ket{\Psi_\mathrm{out}}$, prepared from a specified initial state $\ket{\mathbf{T}}$ and linear transformation $\mathbf{\Lambda}$.  A trial begins with the input state $\ket{\mathbf{T}} = \ket{T_1...T_M} \propto \prod_{i=1}^M (\adagg_i)^{T_i}\ket{0}$, which describes $N = \sum_{i=1}^M T_i $ particles distributed in $M$ input modes in the occupation-number representation. The output state $\ket{\Psi_{\rm out}}$ is generated according to the linear map from output to input mode creation operators $\adagg_i = \sum_{j{=}1}^{M} \Lambda_{ij} \bdagg_j$, where $\mathbf{\Lambda}$ is an $M{\times}M$ matrix.  Finally, the particles in each of the $M$ output modes are counted.  The probability of a particular measurement outcome $\ket{\mathbf{S}} = \ket{S_1...S_{M}}$ is given by
\begin{equation}
	P(\mathbf{S}|\mathbf{T}) = |\langle \mathbf{S} | \Psi_\mathrm{out} \rangle|^2 = \frac{\left | \textrm{Per}(\mathbf{\Lambda}^{(\mathbf{S},\mathbf{T})})  \right | ^2}{\prod_{j=1}^{M} S_j! \prod_{i=1}^M T_i!}
	\label{eq:PST}
\end{equation}
where the $N{\times}N$ submatrix $\boldsymbol\Lambda^{(\mathbf{S},\mathbf{T})}$ is obtained by keeping $S_j$ ($T_i$) copies of the $j^{th}$ column ($i^{th}$ row) of $\boldsymbol\Lambda$\cite{supp}. 

The aim of boson sampling is to sample from the distribution given by $P(\mathbf{S}|\mathbf{T})$. A QBSM achieves this efficiently by directly sampling $|\langle \mathbf{S} | \Psi_{\mathrm{out}} \rangle|^2$. The classical strategy, on the other hand, requires evaluating the right-hand side of Eqn.~\eqref{eq:PST}, for which the calculation time scales exponentially in $N$~\cite{Ryser1963}.

\begin{figure}[h!]
\centerline{\includegraphics[width=1\columnwidth]{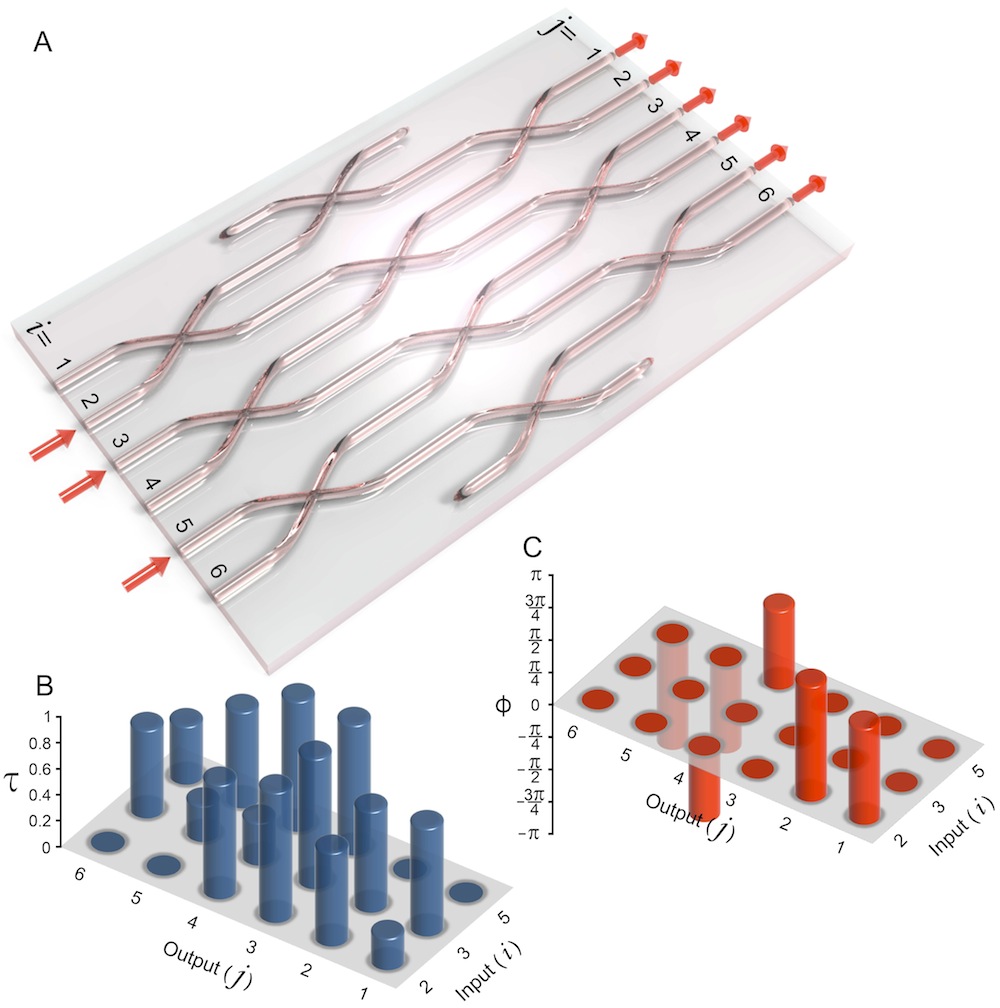}}        
        \caption{\textbf{Schematic and characterization of the photonic circuit.} (a) The silica-on-silicon waveguide circuits consist of $M=6$ accessible spatial modes (labeled $1{-}6$).  For the three-photon experiment, we launch single photons into inputs $i=2$, $3$ and $5$ from two parametric down-conversion sources and postselect outcomes in which three detections are registered amongst the output modes $j$.  For the four-photon experiment, which is implemented on a different chip of identical topology, we inject a double photon pair from a single source into the modes $i=1$, $3$ and postselect on four detection events.  (b)-(c) Measured elements of the linear transformation $\Lambda_{ij}=\tau_{ij}e^{i\phi_{ij}}$ linking the input mode $i$ to the output mode $j$ of our three-photon apparatus. The circuit topology dictates that several $\tau_{ij}$ are zero, and our phase-insensitive input states and detection methods imply only six non-zero $\phi_{ij}$. Since only relative values are needed due to post-selection, we rescale each row of $\tau$ so that its maximum value is unity.}
         \label{fig:chip}
\end{figure}

\begin{figure*}[ht]
	\centerline{\includegraphics[width=1.6\columnwidth]{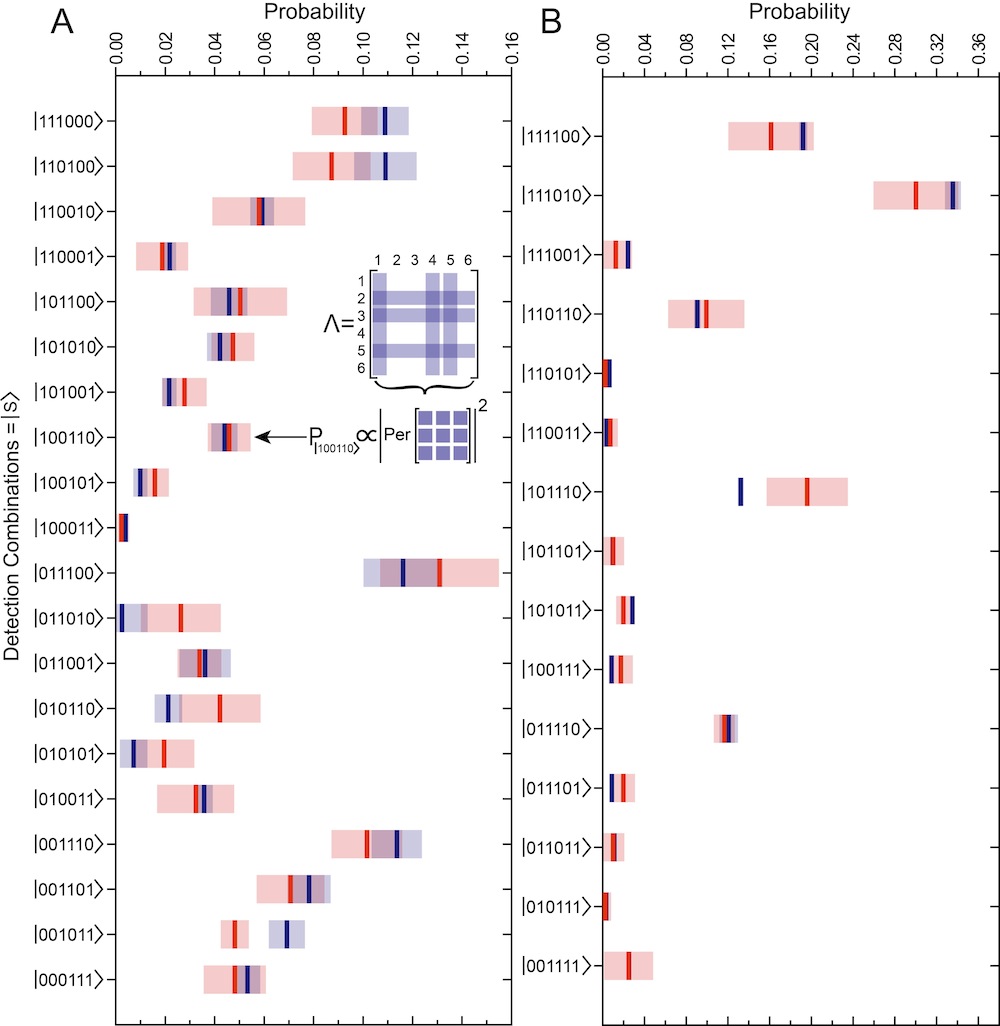}}
	\caption{\textbf{Boson sampling results.} The measured relative frequencies $\mathbf{P}^{\rm exp}$ of outcomes in which the photons are detected in distinct modes are shown in red for (a) three- and (b) four-photon experiments.  Each data set is collected over 160 hours, and statistical variations in counts are shown by the red shaded bars. The theoretical distributions $\mathbf{P}^{\rm th}$ (blue) are obtained from the permanent of a submatrix constructed from the full transformation $\mathbf{\Lambda}$, as depicted in the inset.  The blue error bars arise from uncertainties in the characterization of $\mathbf{\Lambda}$.}
	\label{fig:bosSamp}
\end{figure*}

Our QBSM consists of sources of indistinguishable single photons, a multiport linear optical circuit, and single-photon counting detectors. Two parametric down-conversion (PDC) pair sources \cite{MosleyPJ2008hgu} are used to inject up to four photons into a silica-on-silicon integrated photonic circuit, fabricated by UV writing~\cite{Smith2009}. The circuit is shown in Fig.~\ref{fig:chip}a and consists of $M{=}6$ input and output spatial modes coupled by a network of ten beam splitters~\cite{Metcalf2012}. The output state is measured with single-photon avalanche photodiodes on each mode. We only consider outcomes in which the number of detections equals the intended number of input photons~\cite{supp}.

Our central result of three- and four-boson sampling is shown in Fig.~\ref{fig:bosSamp}. In the first case, we repeatedly inject three photons in the input state $\ket{{\bf T}} = \ket{011010}$, monitor all outputs, and collect all three-fold coincident events. In the four-photon experiment, we use the input $\ket{{\bf T}} = \ket{202000}$ and record all four-fold events.  For each experiment, the measured relative frequencies $P_S^{{\rm exp}}$ for every allowed outcome $\ket{\mathbf{S}}$ are shown along with their observed statistical variation.  The corresponding theoretical $P_S^{{\rm th}}$, calculated using the right-hand side of Eqn.~\eqref{eq:PST}, are shown along with their uncertainties arising from the characterization of $\mathbf{\Lambda}$, described below.

We reconstruct $\mathbf{\Lambda}$ with a series of one- and two- photon transmission measurements to determine its complex-valued elements $\Lambda_{ij}=\tau_{ij}e^{i\phi_{ij}}$  \cite{Laing2012}.  The characterization results for the circuit used in the three-photon experiment are shown in Fig.~\ref{fig:chip}b-c.  To obtain the magnitude $\tau_{ij}$, single photons are injected in mode $i$. The probability of a subsequent detection in mode $j$ is given by $P_1(j,i) =  \left | \Lambda_{ij} \right | ^{2} =  \tau_{ij}^{2}$. The phases $\phi_{ij}$ are determined from two-photon quantum interference measurements. The probability that a photon is detected in each of modes $j_1$ and $j_2$ when they are injected in modes $i_1$ and $i_2$ is given by $P_2(j_1,j_2,i_1,i_2) = \left | \Lambda_{i_1j_1} \Lambda_{i_2j_2} + \Lambda_{i_2j_1} \Lambda_{i_1j_2} \right |^2$. This expression is used to relate the relevant phases $\phi_{ij}$ given the previously determined magnitudes $\tau_{ij}$. We use a least-squares fit to determine the phases $\phi_{ij}$ from a complete set of two-photon measurements~\cite{supp}. 

To analyze the performance of our QBSM we compare our results to an ideal machine. We quantify the match of two sets of relative frequencies $\mathbf{P}^{(1)}$ and $\mathbf{P}^{(2)}$ by calculating the $L_1$ distance $d^{(N)}(\mathbf{P}^{(1)},\mathbf{P}^{(2)}){=}\frac{1}{2}\sum_{S}|P^{(1)}_{S}{-}P^{(2)}_{S}| $, where $N$ denotes the number of photons in a sample~\cite{Gilchrist05}. Identical and maximally dissimilar distributions correspond to $d{=}0$ and $d{=}1$, respectively. For our experiments we calculate $d^{(N)}(\mathbf{P}^{\rm exp},\mathbf{P}^{\rm th})$ to give $d^{(3)}{=}0.094\pm0.014$ and $d^{(4)}{=}0.097\pm0.004$. Even in an ideal QBSM with perfect state preparation and detection, the statistical variations result in nonzero $d$. If we substitute for our experimental data a Monte Carlo sampling of $\mathbf{P}^{{\rm th}}$ with sample size equivalent to our experiments, we instead calculate $d^{(3)}{=}0.043\pm0.012$ and $d^{(4)}{=}0.059\pm0.022$. This suggests there are significant contributions to $d(\mathbf{P}^{{\rm exp}},\mathbf{P}^{{\rm th}})$ beyond statistical deviation.

\begin{figure*}[ht]
\centerline{\includegraphics[width=1.8\columnwidth]{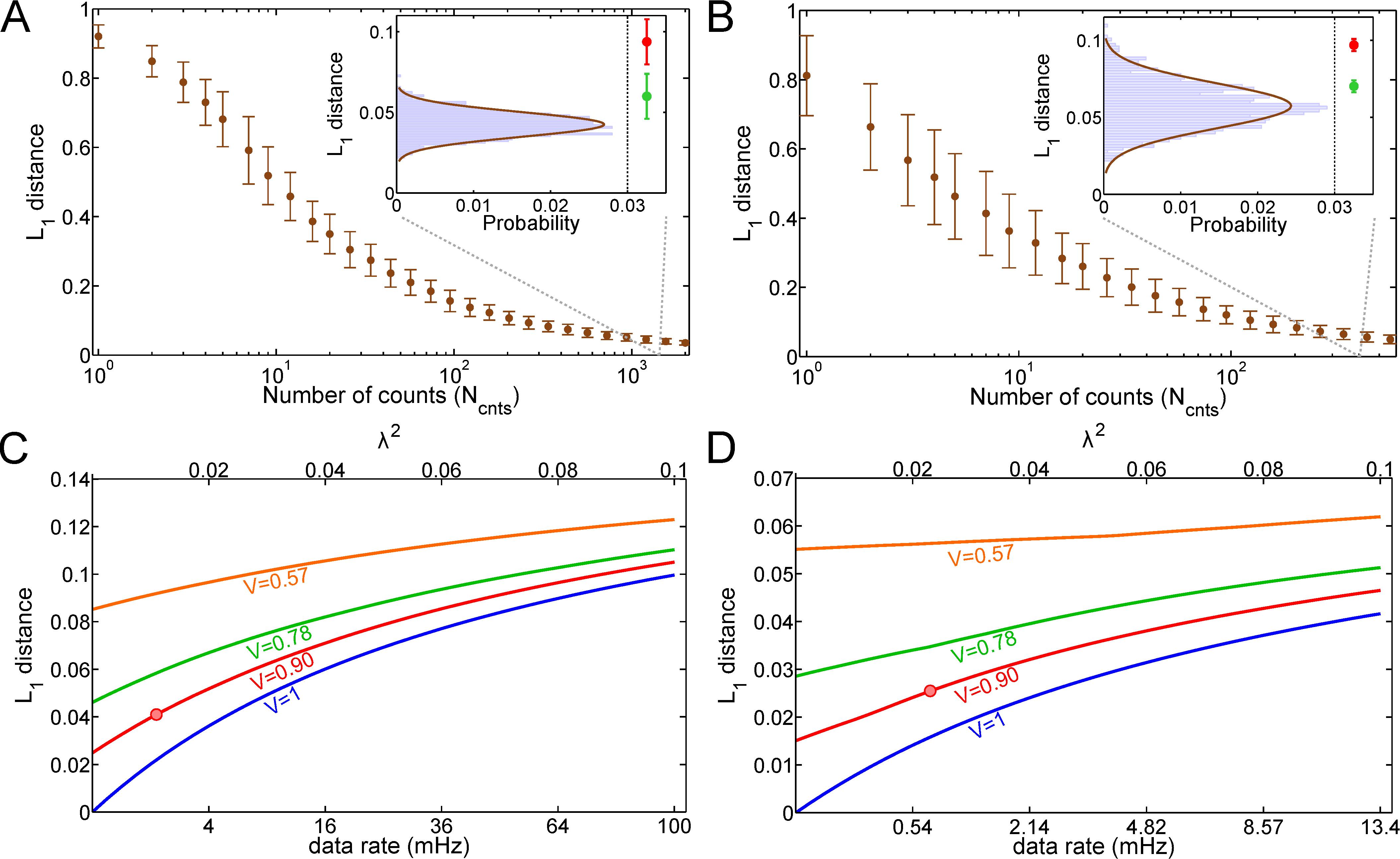}}
	\caption{\textbf{Sampling accuracy.}  We consider several boson outcome distributions: the experimental samples $\mathbf{P}^{\rm{exp}}$, the ideal predictions of Eqn. \eqref{eq:PST} $\mathbf{P}^{\rm{th}}$, and the predictions of the full model $\mathbf{P}^{\rm{mod}}$.  \textbf{(a)} The $L_1$ distance $d$ between $\mathbf{P}^{\rm{th}}$ and a Monte Carlo simulation of an ideal machine that samples $\mathbf{P}^{\rm th}$ a finite number of times for our three and \textbf{(b)} four photon cases.  The inset histograms show the variation in $d$ expected for a sample size corresponding to the $1421$ and $405$ counts collected in our three- and four-photon experiments. The distance $d(\mathbf{P}^{\rm{exp}},\mathbf{P}^{\rm{th}})$ (red) between our experimental sample and the ideal distribution suggests an underlying systematic inaccuracy. A full model which includes multiphoton emission and photon distinguishability is validated by the distance $d(\mathbf{P}^{\rm{exp}},\mathbf{P}^{\rm{mod}})$ (green) which is consistent with statistical variation. \textbf{(c)}  The predicted variation in $d(\mathbf{P}^{\rm th},\mathbf{P}^{\rm mod})$ is shown as a function of $\lambda$ for the photon distinguishabilites $V$ indicated for the three- and \textbf{(d)} four-photon cases. Our experimental conditions are marked (red dot).}
	\label{fig:errors}
\end{figure*}

Due to experimental limitations, our QBSM occasionally samples distributions other than $\mathbf{P}^{\rm th}$. The dominant sources of this sampling inaccuracy are multi-photon emission and partial distinguishability of our photon sources. In practice, all single-photon sources produce multiple photons with a finite probability \cite{Eisa2011}. For our PDC sources, the output state is approximately ${\ket{00}+\lambda\ket{11}+\lambda^2\ket{22}}$, with $\lambda{\ll}1$. Both single-photon and undesired multiphoton terms increase with $\lambda$. In our three-photon experiments, for example, multiphoton emission from the two PDC sources can lead to input states $\ket{\mathbf{T}} = \ket{021010}$ or $\ket{012020}$, which contribute to three-fold coincident events if photons are lost or emerge in the same output mode. In addition, partial distinguishability of the photons contaminates the distribution sampled by the QBSM by mixing in one- and two-photon interference effects\cite{Rohde2012b}. To characterize the photon indistinguishability we performed two-photon Hong-Ou-Mandel experiments, in which interference visibilities were 90$\%$ of the expected values on average. 

We form a new distribution $\mathbf{P}^{\rm mod}$ that accounts for the effects of multiphoton emission and photon distinguishability~\cite{supp}. The distance $d(\mathbf{P}^{\rm exp},\mathbf{P}^{\rm mod})$, shown by the green point in the insets of Fig.~\ref{fig:errors}a-b, is found to be consistent with the statistical variation due to a finite sample size, for both the three- and four-photon experiments. This suggests we have correctly identified and modeled the significant sources of inaccuracy. Our validated full model may be used to predict the performance of realistic QBSMs and guide the design of future machines with larger $N$. To investigate how the performance of our QBSM depends on $\lambda$ and photon distinguishability, we calculate $d(\mathbf{P}^{\rm mod},\mathbf{P}^{\rm th})$ for a range of operating parameters (Fig.~\ref{fig:errors}c-d). In terms of $\lambda$, a clear tradeoff is presented between data rate and inaccuracy due to multiphoton emission, which is an intrinsic consequence of using PDC sources. Improvement in photon indistinguishability increases the fidelity to the ideal machine, and additionally is thought to enhance the computational power of a QBSM~\cite{Rohde2012b}.

Our results demonstrate that boson sampling provides a model of quantum-enhanced computation that is experimentally feasible with existing photonic technology. Future generations of QBSMs will benefit from ongoing advances in integrated photonics such as reduced transmission loss, efficient number-resolving detectors \cite{Gerrits2011}, and multiplexed \cite{Migdall2002, Nunn2012} or single-emitter \cite{Eisa2011} photon sources. Such improvements suggest a promising route to large-scale photonic QBSMs, which will provide clear evidence for the computational power of quantum mechanics.

\textbf{Acknowledgements:} We thank Josh Nunn for valuable insights. This work was supported by the EPSRC(EP/C51933/01 and EP/C013840/1), the EC project Q-ESSENCE (248095), the Royal Society, and the AFOSR EOARD. XMJ and NKL are supported by EU Marie-Curie Fellowships (PIIF-GA-2011-300820 and PIEF-GA-2010-275103).  JBS acknowledges support from the United States Air Force Institute
of Technology. The views expressed in this article are those of the authors and do not reflect the official policy or position of the United States Air Force, Department of Defense, or the U.S. Government.

\bibliographystyle{naturemag}
\bibliography{1231692RevisedBib}

\clearpage


\section{Supplementary Information}\section{Materials and Methods}
\textbf{Multiphoton states generation.}  An $80$ MHz Ti-Sapphire oscillator outputs 100 fs pulses at 830 nm, which undergo type-I second harmonic generation in a $700$ $\mu$m $\beta$-BaB$_2$O$_4$ (BBO) crystal.  This $415$ nm light is then used used to pump type-II collinear parametric downconversion (PDC) in $8$ mm-long AR-coated Potassium Dihydrogen Phosphate (KDP) crystals~\cite{MosleyPJ2008hgu}.  Both modes of the resulting squeezed state are passed through spectral filters (Semrock, $\Delta\lambda{=}3$ nm), to maximize photon indistinguishability. In the three-photon experiment, two such PDC sources are coupled to four polarization-maintaining (PM) fibers~\cite{Metcalf2012}.  One of the photons is sent directly to a heralding detector, while the other three are coupled into the photonic circuit via a PM v-groove array (VGA) on a 6-axis alignment stage at the chip input.  Programmable optical delay is provided by motor-controlled translation stages on each single-photon mode preceding the circuit, allowing the photons to arrive temporally coincident at the interferometric network in Fig 2A. Another PM VGA at the chip output couples the output modes into avalanche photodiode (APD) single photon counting modules (PerkinElmer SPCM-AQ4C), which are monitored by a home-built coincidence counting program loaded onto a commercially available FPGA development board (Xilinx SP605) operating with a 5 ns coincidence window.  In the four-photon experiment, the higher order emission ($\ket{22}$) from a single PDC crystal was launched into two spatial modes of a different chip, though of identical geometry, which had VGAs glued to input and output.  This reduced the coupling efficiency uncertainty for the four-photon QBSM, which reduced the size of the error bars in Fig. 3B. In both cases, to minimize undesired PDC emission terms the sources were pumped as lightly as possible while maintaining reasonable count rates, yielding $\lambda^2{=}0.011$ and $0.023$ in the three- and four-photon sampling experiments respectively.  \\ \\
\textbf{Photonic circuit fabrication.}  The boson sampling is performed on a UV-written silica-on-silicon integrated chip~\cite{Metcalf2012}.  The chip was fabricated by focusing a continuous wave UV laser (244 nm) onto a photosensitive layer to create a local increase in the index of refraction.  The chip is then moved, via computer-controlled precision translation stages, transversely to the incident UV beam to trace out the desired waveguide network geometry.  One can fabricate cross-coupling beamsplitters of a certain reflectivity by crossing waveguides at a specific angle.  Such beamsplitters take up less space than traditional evanescent couplers in such circuits, allowing reduced chip size and lower loss~\cite{Kundys2009}.  While there are eight spatial modes in the middle of the circuit (Fig. 2A), the outer modes are not fabricated to the edge of the chip and are thus not accessible to the experimenter.  They can then be accurately treated as losses on the neighboring modes, leaving a six (accessible) mode interferometric network.  \\ \\
\textbf{Boson sampling data collection.}  For the four-photon QBSM, the FPGA simultaneously counts all possible combinations of coincidence detection events (one detection event, two coincident detections etc) amongst the six APDs  monitoring the output modes of the integrated circuit.  For the three-photon QBSM, a similar set of coincidences is taken, but conditioned on detection of the herald photon by a separate APD.  There is a nonzero background count rate on each detector which contributes to erroneous $N$-fold detection events when $N{-}1$ APDs detect photons and another APD erroneously fires.  The $N{-}1$ coincidence rate is included in the above set of statistics collected by the FPGA and, by temporarily blocking all input modes to the circuit, one can estimate the background rate on each detector.  The resulting background contribution to $N$-fold coincidences, comprising approximately $5\%$ of the total counts, are then subtracted from the raw data, with the results plotted in Fig. 3.    \\ \\
\textbf{Circuit characterization.}  Our photonic chip performs a non-unitary, complex-valued linear mapping of input to output modes $\Lambda_{ij}{=}\tau_{ij} e^{i\phi_{ij}}$. We follow the method outlined in \cite{Laing2012} to use a set of one- and two-photon data to reconstruct $\mathbf{\Lambda}$.  

We can find $\tau_{i_1,j_1}$ by coupling light (single photons) into mode $i_1$ and monitoring the probability of output in $j_1$ yielding
\begin{align}
	\left| \bra{0}\hat{b}_{j_1} \adagg_{i_1}\ket{0} \right |^2 &= \left| \bra{0}\hat{b}_{j_1} \sum_{j=1}^M \Lambda_{i_1j} \bdagg_{j} \ket{0} \right |^2   \nonumber \\
	&= \left| \Lambda_{i_1 j_1} \right |^2 \nonumber \\
	&= \left| \tau_{i_1 j_1} \right |^2 
	\label{eq:1phot}	
\end{align}
It is sufficient to measure the ratio of values between output ports, $\left ( \tau_{i_1,j_1}/\tau_{i_1,j_2} \right )^2$, as this describes the transformation induced by our circuit up to a constant factor for each input, $x_i$.  If $\mathbf{\Lambda}$ is the true transformation, then this procedure gives us $\mathbf{X} \mathbf{\Lambda}$ where $\mathbf{X}$ is a diagonal matrix with entries $X_{ii}{=}x_i$.  However, the properties of the matrix permanent give $\textrm{Per}(\mathbf{X} \, \mathbf{\Lambda}){=}(\prod_i x_i) \textrm{Per}(\mathbf{\Lambda})$.  Therefore, Eq. 1 of the main text shows, since we always launch the same input state, every portion of our boson distribution is multiplied by a constant factor that cancels in the normalization.  This data is collected by periodically pausing boson sampling data collection, blocking two of the three photon inputs, and measuring the relative power in each output mode.  We thus obtain accurate values for $\tau$ as well as a variance that is important for calculating the error bars in Fig. 3.

One can then use two photon interference to find $\phi_{i,j}$ \cite{Laing2012}.  If one inputs two photons into modes $i_1,i_2$ and detect in modes $j_1,j_2$, then we have $\ket{S} {=}\adagg_{j_1}\adagg_{j_2}\ket{0}$ and $\ket{T}{=}\adagg_{i_1} \adagg_{i_2} \ket{0}$.  If single, indistinguishable photons are used, then the probability of post-selecting this output is
\begin{align}
	P_{\rm{indist}} &=  \left | \textrm{Per}(\mathbf{\Lambda^{\mathbf{(S,T)}}}) \right |^2\nonumber \\
		  &= \left |\Lambda^\mathbf{(S,T)}_{11}\Lambda^\mathbf{(S,T)}_{22} + \Lambda^\mathbf{(S,T)}_{12}\Lambda^\mathbf{(S,T)}_{21} \right | ^2 \nonumber \\
		  &= (\tau_{i_1,j_1}\tau_{i_2,j_2})^2 + (\tau_{i_1,j_2}\tau_{i_2,j_1})^2  \nonumber \\
		  & + 2 \tau_{i_1,j_1}\tau_{i_1,j_2}\tau_{i_2,j_1}\tau_{i_2,j_2}  \nonumber \\
		  & \times \cos(\phi_{i_1,j_1}-\phi_{i_1,j_2}-\phi_{i_2,j_1}+\phi_{i_2,j_2})
	\label{eq:indist}	
\end{align}
If the photons launched into the individual modes are distinguishable, then we get the incoherent sum of their individual statistics.  We again take a matrix permanent, but as this is an incoherent process one finds
\begin{align}
  P_{\rm{dist}} &=(\tau_{i_1,j_1}\tau_{i_2,j_2})^2 + (\tau_{i_1,j_2}\tau_{i_2,j_1})^2
  \label{eq:dist}	
\end{align}
One can then perform a Hong-Ou-Mandel experiment and find that the resulting interference visibility is
\begin{align}
	V &= \: \frac{P_{dist}-P_{indist}}{P_{dist}} \nonumber \\
	  &= \: \frac{2\tau_{i_1,j_1}\tau_{i_1,j_2}\tau_{i_2,j_1}\tau_{i_2,j_2}}{(\tau_{i_1,j_1}\tau_{i_2,j_2})^2 + (\tau_{i_1,j_2}\tau_{i_2,j_1})^2} \nonumber \\
	  & \times \cos(\phi_{i_1,j_1}+\phi_{i_2,j_2}-\phi_{i_1,j_2}-\phi_{i_2,j_1})
	  \label{eq:HOMvis}
\end{align}
We perform this two-photon interference experiment and fit a Gaussian to the resulting data to determine the visibility.  Using the known $\tau_{i,j}$, one can then find $|\phi_{i_1,j_1}+\phi_{i_2,j_2}-\phi_{i_1,j_2}-\phi_{i_2,j_1}|$.  Repeating this procedure for all accessible two photon dips, and applying additional constraints one can determine $\phi_{i,j}$ \cite{Laing2012}.  


For our circuit geometry, we encounter an overconstrained problem as we measure more interference visibilities than unknown $\phi$.  Therefore, we run a least squares minimization to find the set of $\phi$ that best fits our two photon interference data.  To find the error bars in Fig. 3, we use a Monte Carlo method where the elements of $\mathbf{\Lambda}$ are selected from a normal distribution with an appropriate variance for each element.   The one photon measurement was repeated periodically during the $160$ hour long boson sampling data collection, thus yielding the variance in the $\tau_{ij}$ characterization, which was determined to dominate the uncertainty in the predicted boson distribution in Fig. 3.  This explains why the four-photon QBSM, which had VGAs glued to the ends and thus was much less susceptible to changes in the coupling, has significantly smaller error bars in the predicted distribution shown in Fig. 3B.  The uncertainty in the Gaussian fit to the two-photon interference patterns and in $\tau_{ij}$ was then used to determine the variance in $\phi_{ij}$. 

This characterization process is efficient, as a general linear transformation over $M$ modes can be described by $\orderof{(M^2)}$ parameters, requiring $\orderof{(M^2)}$ measurements with this technique.  While photons from the PDC sources were used for the circuit characterization here, one can also use classical coherent states~\cite{Laing2012,Rahimi-Keshari2012}.  However, the single-photon based technique outlined here benefits from not requiring phase-stable path length matching and, because we use the same sources for boson sampling and characterization, the photonic degrees of freedom (polarization, spectrum etc) for the characterization match that used in the experiment.

With $\Lambda_{i,j}$ experimentally determined over the accessible modes, one can predict the post-selected boson distribution for any input/output $\ket{S}$, $\ket{T}$ by constructing $\mathbf{\Lambda}^\mathbf{(S,T)}$ from $\mathbf{\Lambda}$ and taking the permanent according to Eq. (1) in the main text.  Thus, we effectively use the one and two photon boson distributions to characterize our non-unitary operation over the accessible modes.  One can then take various matrix permanents of this non-unitary matrix to predict the boson distribution for \textit{any} $N$.

\section{Supplementary Text}

In this section we first outline how the boson distribution is given by a set of matrix permanents.  We then show how the boson distribution can be accurately predicted by the permanents associated with a non-unitary matrix describing a lossy channel, $\mathbf\Lambda$, if one post-selects on trials where no photons are lost.  Finally, the principal sources of error in this experiment, namely the photon distinguishability and higher order terms from our PDC photon sources, are discussed.

\subsection{The boson distribution is given by a set of matrix permanents}
We assume $N$ bosons are injected into a network that performs a unitary transformation over $M$ modes.  We consider the special case, appropriate to our experiment, where the input (and output) states contain no more than one boson per mode, though the general case is treated elsewhere~\cite{Scheel2004}.  Without loss of generality, let modes $1$ to $N$ contain an input boson, while modes $N{+}1$ to $M$ have vacuum inputs.  The input state can then be described by

\begin{equation}
	\ket{\Psi_{\rm in}} = \ket{T} =  \prod_{i=1}^N  \adagg_{i} \ket{0}  
\end{equation}
The unitary transformation allows one to evolve the operators according to
\begin{equation}
	\adagg_i = \sum_{j=1}^M U_{ij} \bdagg_j
	\label{eq:Uevol}
\end{equation}
where $\adagg_i$ and $\bdagg_j$ are creation operators on the $i$-th input and $j$-th output mode respectively.  We then obtain the output state
\begin{equation}
	\ket{\Psi_{\rm out}} = \prod_{i=1}^N \left ( \sum_{j=1}^M U_{ij} \bdagg_j \right ) \ket{0}
\end{equation}

To find the boson distribution, we project our output onto a state $\ket{\mathbf{S}}$ which, in the number state basis, we describe by an $N$ element vector $\mathbf{S}$, where $S_j$ gives the mode of the $j$-th boson.  The probability of measuring this state is then
\begin{align}
	\label{eq:PTfullS}
	 P_S &= |\ip{\mathbf{S}}{\psi_{\rm out}}|^2		\\
	     &= \left | \bra{0} \prod_{i=1}^N \hat{b}_{S_i} \left[ \prod_{j=1}^N \left ( \sum_{k=1}^M  U_{jk} \bdagg_{k} \right ) \right] \ket{0} \right | ^2 \nonumber
\end{align}

The term in square brackets can be expanded and includes $M^N$ terms, as one is selecting $N$ bosons from $M$ modes where repetitions are allowed ($>1$ boson in a mode).  One can rewrite this term in square brackets to give
\begin{equation}
	P_S = \left | \bra{0} \prod_{i=1}^N \hat{b}_{S_i} \left[ \sum_{j=1}^{M^N} \left ( \prod_{k=1}^N U_{k,\widetilde{V}_k^j} \bdagg_{\widetilde{V}_k^j} \right ) \right] \ket{0} \right |^2
	\label{eq:PTwithV}
\end{equation}
where $\mathbf{\widetilde{V}}$ is the set of $M^N$ permutations of $N$ photons amongst $M$ modes, repetitions allowed.  The tilde notation will be used throughout this paper for a set of permutations.  Then, $\widetilde{V}_k^j$ indicates the mode of the $k$-th boson in the $j$-th permutation.  As an example, consider the case with $M=3$ modes and $N=2$ input bosons, then
\begin{equation}
\begin{matrix}
	\widetilde{V}^1 = [1,1]  &	\widetilde{V}^2 = [1,2]  &	\widetilde{V}^3 = [1,3]  \\
	\widetilde{V}^4 = [2,1]  &	\widetilde{V}^5 = [2,2]  &	\widetilde{V}^6 = [2,3]  \\
	\widetilde{V}^7 = [3,1]  &	\widetilde{V}^8 = [3,2]  &	\widetilde{V}^9 = [3,3]
\end{matrix}
\end{equation}
Let us denote all $N!$ permutations of $\mathbf{S}$ by $\mathbf{\widetilde{S}}$ where $\widetilde{S}_k^{j}$ indicates the mode of the $k$-th boson in the $j$-th permutation.  For example, if we project onto the state $\ket{S}{=}\ket{011}$ then $\mathbf{S}{=}[2,3]$ and
\begin{equation}
\begin{matrix}
	\widetilde{S}^1 = [2,3]  &  \widetilde{S}^2 = [3,2] 
\end{matrix}
\end{equation}
It is clear we only retain terms from the summation in Eq.~\ref{eq:PTwithV} where $\widetilde{V}^j\in\widetilde{S}$, otherwise at least one annihilation operator will act on vacuum and give $P_S{=}0$.  This then leaves us with
\begin{equation}
	P_S = \left | \sum_{j=1}^{N!}  \prod_{k=1}^N U_{k,\widetilde{S}_k^j}  \right |^2
	\label{eq:PTtildeT}
\end{equation}
The formula for the permanent of an $n \times n$ matrix $A$ with elements $a_{ij}$ is
\begin{equation}
	\textrm{Per}(A) = \sum_{i=1}^{n!} \prod_{j=1}^n a_{j,\widetilde{\sigma}_j^i}
	\label{eq:perm}
\end{equation}
where $\widetilde{\sigma}_j^i$ gives the $j$-th element of the $i$-th permutation of the numbers $1,2,....,n$.  The term inside the modulus of Eq.~\ref{eq:PTtildeT} has the same form as the matrix permanent in Eq.~\ref{eq:perm}.  Our original unitary, $U$, can be described by an $M \times M$ matrix.  However, it is obvious from Eq.~\ref{eq:PTtildeT} that, in general, we take the permanent of a subsection of $U$.  Specifically, we only keep rows $1 \rightarrow N$, those rows corresponding to modes with input photons.  In addition, we only keep columns corresponding to the elements of $S$.  Let us call this modified subsection of our original unitary $U^{(\mathbf{S,T})}$.  Then, using the definition of the permanent we can rewrite Eq.~\ref{eq:PTtildeT}
\begin{align}
	P(\mathbf{S}|\mathbf{T}) &= \left | \textrm{Per}(U^\mathbf{(S,T)}) \right |^2
	\label{eq:Ust}
\end{align}
If one allows the possibility of more than one photon per input and output mode, then a similar analysis yields Eq. 1 in the main text~\cite{Scheel2004}.  

We also note that the above treatment assumes the bosonic commutation relation $[\bdagg_i,\bdagg_j]=0 \:\forall\: i,j$.  If the system consisted of indistinguishable fermions, then the corresponding anticommutation relation $\{\bdagg_i,\bdagg_j\}=0 \:\forall\: i,j$ would be used, leading to alternating plus and minus signs introduced in the summation in Eq. \ref{eq:PTtildeT}, yielding the easily classically computable determinant.

\subsection{Effects of loss}
In any experimental implementation of boson sampling there will be losses.  These losses, regardless of where they occur, can be modeled as beam splitters that link accessible to inaccessible modes~\cite{Thomas-Peter2011a}.  When these losses are considered, it is important to ask whether the boson distribution is still given by a set of matrix permanents and, if so, what linear transformation does that matrix describe.

Let us adopt the convention that modes $1$ to $M$ are accessible modes while inaccessible loss modes are given the labels $M+1$ to $L$.  There is an $L \times L$ unitary operation describing the evolution of our pure input state, though we must trace over these loss modes at the output, yielding a mixed state over the accessible modes.  Let us assume that photons are input into modes $1$ to $N$ where $N\leq M$ and we post-select on cases where $N$ photons are detected, by definition, in the accessible modes $1$ to $M$.  Then Eq.~\ref{eq:PTwithV} becomes
\begin{equation}
	P_S = \left | \bra{0} \prod_{i=1}^N \hat{b}_{S_i} \left[ \sum_{j=1}^{L^N} \left ( \prod_{k=1}^N U_{k,\widetilde{V}_k^j} \bdagg_{\widetilde{V}_k^j} \right ) \right] \ket{0} \right | ^2 
	\label{eq:PTloss}
\end{equation}
but $S_i \leq M$ as we can only project on accessible modes.  Therefore, even though $U$ is an $L \times L$ matrix and the elements of $\widetilde{V}$ range from $1\rightarrow L$, when we project onto $\widetilde{S}$ (all the permutations of $S$) we are left with
\begin{align}
	P_S &= \left | \sum_{j=1}^{N!}  \prod_{k=1}^N U_{k,\widetilde{S}_k^j}  \right |^2 \nonumber \\
	 		&= \left | \textrm{Per}(\mathbf{U}^\mathbf{(S,T)}) \right |^2 = \left | \textrm{Per}(\mathbf{\Lambda}^\mathbf{(S,T)}) \right |^2
	\label{eq:UstLoss}
\end{align}
where $U^\mathbf{(S,T)}$ is again a modified version of the original unitary but only keeping rows $1\rightarrow N$ and columns in $S_i' \leq M$.  Since these elements always describe the accessible modes, then we can equivalently work in terms of $\Lambda$, where $\Lambda_{i,j}{=}U_{i,j}$ but $i,j \leq M$.  In summary, when post-selecting on no bosons being lost, one can work in terms of $\Lambda$, a non-unitary linear transformation that is simply the subsection of $U$ over the accessible modes.  The matrix permanents of such a non-unitary linear transformation lead to the theoretical predictions in Fig. 3 of the main text.

Equivalently, we can describe our system as a noisy (lossy) quantum channel in the operator sum representation \cite{Nielson04}.  This formalism will be useful later when we discuss sources of error from higher order PDC terms.  In this picture, the accessible modes are in the system $Q$, while all inaccessible loss modes form the environment system $E$.  We can describe the transformation induced by our circuit over the full space $Q \otimes E$ by a unitary operation $U$.  Let $\rho$ and $\sigma$ be the inputs to $Q$ and $E$ respectively, then the output in $Q$ after a projective measurement $P_m$ and tracing over the environment is described by
\begin{equation}
	\rho_{\rm{out}} = \textrm{tr}_E({P_m U(\rho \otimes \sigma) U^\dag P_m})
	\label{eq:emRho}
\end{equation}
Let the basis for $E$ be described by $\ket{e_k}$ and the initial state of the environment be $\sigma=\sum_j q_j \ket{j}\bra{j}$, then we can express Eq.~\ref{eq:emRho} as
\begin{equation}
	\rho_{\textrm{out}} = \sum_{jk} E_{jk}\rho E_{jk}^\dag
	\label{eq:Ejk}
\end{equation}
where $E_{jk}=\sqrt{q_j}\bra{e_k}P_m U \ket{j}$ are the Kraus operators.  We do not directly characterize $U$, as it extends over the environment which is inaccessible to the experimenter.  However, with photons we can assume $\sigma = \ket{0}\bra{0}$, and all of our boson sampling results post-select on the case where no photons are lost to the environment.  Therefore, the summation in Eq.~\ref{eq:Ejk} reduces to only one term with postselection
\begin{equation}
	\rho_{\rm{out}} = E_{00} \rho E_{00}^\dag
	\label{eq:E00}
\end{equation}
Experimental boson sampling efforts will sample a non-unitary transformation that is equivalent, when post-selecting on no bosons being lost, to the $E_{00}$ Kraus operator.  

\subsection{Sources of error}
The computational difficulty for a classical machine to sample a boson distribution increases as the maximum error threshold is lowered.  Therefore, while a QBSM need not sample the true boson distribution perfectly~\cite{Aaronson}, it will be easier to beat a classical machine if future QBSMs designs minimize their sampling errors.  In this paper, we have benchmarked the accuracy of our QBSMs by inferring the probability distribution from our data, labeled $\rm{\mathbf{P}}^{\rm{exp}}$, and comparing it to the distribution obtained from Eq. 1, labeled $\rm{\mathbf{P}}^{\rm{th}}$.  Throughout the text, we quantify the distance between two probability distributions via the $L_1$ distance,  $d(\rm{\mathbf{P}}^{(1)},\rm{\mathbf{P}}^{(2)}){=}\frac{1}{2}\sum_i|\rm{P}^{(1)}_i-\rm{P}^{(2)}_i|$.

Our method of benchmarking QBSM accuracy will always yield a nonzero $d$ due to the finite number of collected samples.  We perform a Monte Carlo simulation of $\mathbf{P}^{\rm{exp}}$ from a QBSM that perfectly samples $\mathbf{P}^{\rm{th}}$ as a function of the number of counts collected (Fig. 4, A and B), to show the rate at which $d$ asymptotically approaches zero as the number of samples collected increases.  In the inset histograms, we show the range of outputs for this ideal QBSM for the actual number of experimental counts collected in the three and four photon cases, while the red dots indicate $d(\mathbf{P}^{\rm{th}},\mathbf{P}^{\rm{exp}})$.  These two probability distributions show close agreement in Fig. 3, however a comparison of the histogram and red dots in Fig. 4, A and B indicates there is an additional source of error beyond the finite number of samples.  

Due to experimental limitations, occasionally we sample distributions other than $\mathbf{P}^{\rm{th}}$.  We model the effect of two such imperfections, photon distinguishability and higher order terms from our PDC sources, and form a new distribution $\mathbf{P}^{\rm{mod}}$ that accounts for these effects.  We ignore the effects of photon impurity, as our post-selected data collection and use of nearly-spectrally factorable photon sources~\cite{MosleyPJ2008hgu}, minimizes this contribution.  We find that the new $d(\mathbf{P}^{\rm{exp}},\mathbf{P}^{\rm{mod}})$, indicated with the green dot in Fig. 4, A and B, comes well within the output variance of an ideal machine.  This indicates we have correctly diagnosed and modeled the principal sources of experimental error, which will be important in guiding designs of future, larger $N$, QBSMs.

\subsubsection{Effect of using heralded single photon sources}
The parametric downconversion sources we use to generate our photons actually generate a two-mode squeezed state that is given by
\begin{equation}
	\ket{\Psi_{\rm{PDC}}} = \sqrt{1-\lambda^2}\sum_{n=0}^{\infty}\lambda^n \ket{nn}
	\label{eq:PDCstate}
\end{equation}
where $0\leq \lambda < 1$ is the squeezing parameter whose magnitude is determined, in part, by the type of crystal and pump power used.  For our experiment we wish to minimize higher order terms ($\ket{22}$ and $\ket{33}$ for the three- and four-photon QBSMs respectively) and so we deliberately lower our pump power as much as possible while maintaining a feasible count rate.  For the three-photon experiment, $\lambda{=}\sqrt{0.011}$ and for the four-photon experiment $\lambda{=}\sqrt{0.023}$.  However, even in this case we will sometimes inject more than $N$ photons into our circuit which, due to losses, could be observed as an $N$-fold detection at the output.

Due to the circuit characterization method employed, we have no information about what such terms will be.  In the operator sum representation, our characterization only determines the $E_{00}$ Kraus operator, which describes the transformation of our input state when the environment, which includes all loss modes, starts and ends with zero photons.  For example, if we instead inject five photons and lose two, then this process is described by the $E_{20}$ operator, about which we have no information.

To model the effect of using such squeezed sources, we start with a circuit with the same geometry used in the experiment.  Non-uniform losses throughout the circuit can then be modeled by adding beam splitters that link the depicted accessible modes shown in Fig. 2A to inaccessible loss modes, where the beam splitter reflectivity indicates the loss in that channel~\cite{Thomas-Peter2011a}.  Such `loss beamsplitters' are added throughout the circuit.  We cannot directly characterize these losses in our current circuit, however we can estimate them numerically.  The characterized linear transformation $\mathbf{\Lambda}$ is a function of these losses as well as the fabricated interferometers shown in Fig. 2A.  We have performed a loss-independent characterization of the interferometers \cite{Metcalf2012}, and then input this data into a genetic algorithm to find the relative losses between modes that best reproduces the characterized $\mathbf{\Lambda}$.  We then apply three independent scaling factors to the relative losses at the sources, circuit and detectors.  The source scaling factor is chosen to match the known source heralding efficiency, which is a measure of the loss in each source arm.  The detector losses are scaled such that no detector has an efficiency greater than $50\%$, which is appropriate for APDs detecting photons at $830$ nm.  Finally, we scale the relative losses in the circuit to reproduce the known overall system transmission observed experimentally.

With knowledge of these losses, we reproduce an actual unitary linear transformation $\mathbf{U}$ that extends over both the $N{=}6$ modes as well as all loss modes and can be used to accurately predict the effect of higher order PDC terms.  Taking the three-photon QBSM as an example, we use Eq. 1 of the main text to find the probability distributions when the input is $\ket{\mathbf{T}}{=}\ket{011010}$ (desired single-photon input), as well as $\ket{011020}$ or $\ket{022010}$ (the first higher-order terms from our two sources), which we label $\mathbf{P}^{\rm{111}}$, $\mathbf{P}^{\rm{112}}$ and $\mathbf{P}^{\rm{221}}$ respectively.  The boson distributions are then found by summing the terms where three photons appear in the desired accessible modes and zero, one, or two (respectively) photons appear in any combination of loss modes.  For example, the probability of obtaining $\ket{\mathbf{S}}{=}\ket{111000}$ given input $\ket{\mathbf{T}}=\ket{022010}$ is the summation of all terms where three photons appear in the first three accessible modes and two photons appear in any combination of loss modes.  The higher order probability distributions, $\mathbf{P}^{\rm{221}}$ and $\mathbf{P}^{\rm{112}}$, are weighted by $\lambda^2$ which is obtained via a conditional second order correlation measurement, $g^{(2)}(0)$\cite{SmithBJ2009ppg}.  As $\lambda^2$ is small, we only consider the first higher order terms from each source.

\subsubsection{Photon Distinguishability}
Boson sampling assumes indistinguishable bosons, while experimental implementations will always have some distinguishability.  Assuming pure inputs we follow the notation of \cite{Rohde2012b} to write our input state as
\begin{equation}
	\ket{\psi_{\rm{in}}}=\prod_i^N \left (\alpha A^\dagger_{\xi_0,i} + \sqrt{1-\alpha^2} A^\dagger_{\xi_i,i} \right) \ket{0}
	\label{eq:distInput}
\end{equation}
where $N$ is again the number of photons, $\alpha$ is a distinguishability parameter, and $A^\dagger_{\xi_j,i}$ is the creation operator for photon $i$ in mode $\xi_j$.  Each photon is in a superposition of a desired mode $\xi_0$ and another mode $\xi_i$.  By analyzing the reduction in HOM dip visibility at a beamsplitter inside our circuit we find $\alpha = 0.974$ on average in our experiment.

If one photon is distinguishable from the others, then the new probability distribution is given by the permanents of $N{-}1$ matrices which are incoherently summed.  For example, assume an input state $\ket{\mathbf{T}}{=}\ket{1}^{\tau}\ket{11000}$ where $\tau$ labels a distinguishable photon, then for a unitary transformation $U$ the probability of obtaining an output $\ket{\mathbf{S}}$ is
\begin{align}
	P(\mathbf{S}|\mathbf{T}) &= \left |U^\mathbf{(S,T)}_{11}\right|^2 |U^\mathbf{(S,T)}_{22}U^\mathbf{(S,T)}_{33} + U^\mathbf{(S,T)}_{23}U^\mathbf{(S,T)}_{32}|^2 \nonumber \\ 
	       &  + \left |U^\mathbf{(S,T)}_{12} \right|^2 \left| U^\mathbf{(S,T)}_{21}U^\mathbf{(S,T)}_{33} + U^\mathbf{(S,T)}_{23}U^\mathbf{(S,T)}_{31}\right|^2 \nonumber \\
	       &  + \left |U^\mathbf{(S,T)}_{13} \right|^2 \left| U^\mathbf{(S,T)}_{21}U^\mathbf{(S,T)}_{32} + U^\mathbf{(S,T)}_{22}U^\mathbf{(S,T)}_{31}\right|^2 
	\label{eq:PSTdist}
\end{align}
where the terms in parentheses are permanents of $2\times2$ matrices.  We calculate these probability distributions when one photon is distinguishable and weight them by $|\alpha^2\sqrt{1-\alpha^2}|^2$ and $|\alpha^3\sqrt{1-\alpha^2}|^2$, the probability that one photon is distinguishable from the others for the three- and four-photon cases respectively.  As $\alpha$ is large, we ignore the case when two photons are distinguishable.
\end{document}